# The Valley Hall Effect in MoS$_2$ Transistors

Kin Fai Mak[1,2]†, Kathryn L. McGill[2], Jiwoong Park[1,3], and Paul L. McEuen[1,2]*
1. Kavli Institute at Cornell for Nanoscale Science, Ithaca, NY 14853, USA
2. Laboratory of Atomic and Solid State Physics, Cornell University
3. Department of Chemistry and Chemical Biology, Cornell University

Correspondence to: *plm23@cornell.edu, †km627@cornell.edu

**Abstract**:
Electrons in 2-dimensional crystals with a honeycomb lattice structure possess a new valley degree of freedom (DOF) in addition to charge and spin. Each valley is predicted to exhibit a Hall effect in the absence of a magnetic field whose sign depends on the valley index, but to date this effect has not been observed. Here we report the first observation of this new valley Hall effect (VHE). Monolayer MoS$_2$ transistors are illuminated by circularly polarized light which preferentially excites electrons into a specific valley, and a finite anomalous Hall voltage is observed whose sign is controlled by the helicity of the light. Its magnitude is consistent with theoretical predictions of the VHE, and no anomalous Hall effect is observed in bilayer devices due to the restoration of crystal inversion symmetry. Our observation of VHE opens up new possibilities for using the valley DOF as an information carrier in next-generation electronics and optoelectronics.

The charge and spin degrees of freedom (DOF) of electrons are at the heart of modern electronics. They form the basis for a wide range of applications such as transistors, photodetectors and magnetic memory devices. Interestingly, electrons in 2-dimensional (2D) crystals that have a honeycomb lattice structure possess an extra valley DOF *(1)* in addition to charge and spin. This new DOF has the potential to be used as an information carrier in next-generation electronics *(2-6)*. Valley-dependent electronics and optoelectronics based on semimetallic graphene, a representative 2D crystal, have been theoretically proposed *(2-5)*, but the presence of inversion symmetry in the crystal structure of pristine graphene makes both optical and electrical control of the valley DOF very difficult.

In contrast, monolayer molybdenum disulfide (MoS$_2$), a 2D direct band gap semiconductor *(7, 8)* that possesses a staggered honeycomb lattice structure, is inversion asymmetric. Its fundamental direct energy gaps are located at the K and K' valleys of the Brillouin zone as illustrated in figure 1A. Due to the broken inversion symmetry in its crystal structure, electrons in the two valleys experience effective magnetic fields (proportional to the Berry curvature *(4)*) with equal magnitudes but opposite signs (figure 1A). Such a magnetic field not only defines the optical selection rules *(6)* that allow optical pumping of valley-polarized carriers by circularly polarized photons *(9-13)*, but also generates an anomalous velocity for the charge carriers *(6, 14)*. Namely, when the semiconductor channel is biased, electrons from different valleys move in opposite directions perpendicular to the drift current, a phenomenon called the valley Hall effect (VHE) *(4-6, 15)*. The VHE originates from the coupling of the valley DOF to the orbital motion of electrons *(4, 9)*. This is closely analogous to the spin Hall effect (SHE) *(16-20)* with the spin-polarized electrons replaced by valley-polarized carriers.

Under time reversal symmetry, equal amounts of Hall current from each valley flow in opposite directions so that no net Hall voltage is produced. To measure the valley Hall effect, we explicitly break time reversal symmetry by shining circularly polarized light onto a Hall bar device as shown in figure 1B. A population imbalance between the two valleys (i.e. a valley polarization) is thus created. Under a finite bias, both photoconduction (associated with the normal drift current of the photoexcited charge carriers) and a net transverse Hall voltage (associated with the VHE) should occur *(5, 6)*. The presence of a photoinduced anomalous Hall effect (AHE) driven by a net valley polarization is the experimental manifestation of the VHE in monolayer $MoS_2$.

Furthermore, the magnitude of the AHE can be quantified by an anomalous Hall conductivity $\sigma_H$. In general both the intrinsic Berry curvature effect and extrinsic effects from disorder-induced scattering can contribute to $\sigma_H$ *(6, 21)*. Including the intrinsic effect and the side-jump contribution *(4, 6)*, the absolute value of $\sigma_H$ can be written in the simple form (see Supplementary Materials for a derivation)

$$|\sigma_H| \approx \frac{\hbar^2 \pi \Delta n_v}{2 m_e E_g} \frac{e^2}{h}. \qquad (1)$$

Note that $\sigma_H$ is linear in $\Delta n_v$, the carrier density imbalance between the two valleys generated by photoexcitation. Here $m_e \approx 0.4 m_0$ is the electron band mass *(22)* ($m_0$ is the free electron mass) and $E_g \approx 1.9$ eV is the band gap of monolayer $MoS_2$ *(7)*. Equation 1 thus allows for a quantitative comparison between experiment and theory. Note that we only need to consider the density of the majority carriers, which are electrons in our devices (see below).

Figure 1C shows the gate ($V_g$) dependence of the conductivity of a monolayer $MoS_2$ device ($\sigma_{xx}$) extracted from 2-point and 4-point measurements. Unless otherwise indicated, all measurements were performed on monolayer $MoS_2$ at 77 K (see Supplementary Materials for measurement technique and device fabrication details). The usual n-type field effect transistor behavior is seen *(23)*. We also see that the 2-point (measured at $V_x = 0.5$ V) and 4-point conductivities are similar in magnitude, reflecting the presence of near-Ohmic contacts in our device *(24)*. This is further illustrated by the inset, which shows the bias ($V_x$) dependence of the longitudinal current ($I_x$) at different gate voltages $V_g$. Although the $I_x$-$V_x$ characteristic shows the presence of Schottky barriers at small bias, it has no significant influence on our measurements at high bias. A 4(2)-point carrier mobility of 98(61) $cm^2$ $V^{-1}$ $s^{-1}$ is extracted at high $V_g$, where the $\sigma_{xx}$-$V_g$ dependence becomes linear (see Supplementary Materials for temperature-dependent electrical transport).

In figure 1D we examine the photoresponse of our device: this allows us to identify the appropriate photon energy ($E$) for efficient injection of valley-polarized carriers *(10, 25)*. The inset shows the photocurrent $\Delta I_x$ as a function of $V_x$ (at $V_g = 0$ V) under different laser excitation intensities $P$. The data was taken with a focused laser beam (wavelength centered at 657 nm) located at the center of the device. Similar to the effect of electrical gating (see inset of figure 1C), the effect of incident photons is to increase the channel conductivity $\sigma_{xx}$, which indicates that photoconduction is the main mechanism driving the photoresponse in our device *(26)* (see Supplementary Materials for details). The change in conductivity with and without laser illumination $\Delta \sigma_{xx} \equiv \sigma_{xx,light} - \sigma_{xx,dark}$ as a function of incident photon energies $E$ is shown in figure 1D. It clearly shows the A (at $E \approx 1.9$ eV) and B (at $E \approx 2.1$ eV) resonances of monolayer $MoS_2$ *(7)*.

By parking the laser spot at the center of the device, we study the Hall response under on-resonance excitation (wavelength centered at 657 nm, $E \approx 1.89$ eV). To enhance our detection sensitivity, we modulate the polarization state of the incident light at 50 kHz by use of a photoelastic modulator, and we measure the anomalous Hall voltage $V_H$ with a lock-in amplifier (see Supplementary Materials). Under quarter-wave modulation (i.e. $\Delta\lambda = 1/4$), the degree of excitation ellipticity can be continuously varied by changing $\theta$, the angle of incidence of the linearly polarized light with respect to the fast axis of the modulator. On the other hand, half-wave modulation (i.e. $\Delta\lambda = 1/2$) allows us to modulate linear excitations between $-\theta$ and $\theta$. To indicate the special case of quarter-wave modulation with $\theta = 45^0(-45^0)$, in which the polarization is modulated from right-(left-) to left-(right-) handed, we use the notation R-L(L-R) below.

In figure 2A we show the bias $V_x$-dependence of the anomalous Hall voltage ($V_H$) at $V_g = 0$ V (see Supplementary Materials for scanning photocurrent and Hall voltage images). A small but finite $V_H$ that scales linearly with $V_x$ is observed under R-L modulation (solid red line). This is the signature of a photoinduced AHE driven by a net valley polarization. The sign of the signal is reversed when the excitation is changed to L-R modulation (dashed red line). In contrast, no net Hall voltage is seen when we switch to a linear (s-p) modulation (dotted red line, see Supplementary Materials for measurements on other monolayer devices).

To study the polarization dependence carefully, the anomalous Hall resistance $R_H = V_H/I_x$ as a function of the angle $\theta$ is shown in figure 2B for both the quarter- and half-wave modulations. We see that the Hall resistance $R_H$ exhibits a sine dependence on $\theta$ under quarter-wave modulation. A maximum Hall resistance of about 2 Ω is measured under an excitation intensity of ~150 μW μm$^{-2}$. For comparison, zero Hall resistance is observed under half-wave modulation. Our results are consistent with recent experimental observations of a net valley polarization under the optical excitation of the A resonance with circularly polarized light *(9-13)*. The sine dependence of the quarter-wave modulation data reveals the linear relationship between the degree of valley polarization and the excitation ellipticity *(5, 6)*. Specifically, no net valley polarization is generated under linearly polarized excitations.

The possible existence of the photoinduced AHE in bilayer MoS$_2$ devices is investigated under on-resonance excitation and is shown in figure 2A. No noticeable Hall voltage under R-L modulation (as well as under L-R) is observed (solid blue line). The absence of the AHE is further illustrated in figure 2B. The Hall resistance in the bilayer device is nearly zero and is independent of $\theta$ (solid blue dots). The stark contrast between mono- and bilayer devices therefore suggests that an intervalley population imbalance is required to drive the AHE. No such imbalance can be produced in bilayer MoS$_2$ *(4-6)* due to the restoration of inversion symmetry in the crystal structure *(10, 13)* (The role of spin-orbit coupling and of the SHE will be discussed below).

Finally, the effect of the intervalley relaxation of excited carriers on the AHE in MoS$_2$ monolayers is studied in figure 2C. The dependences of the anomalous Hall conductivity $\sigma_H = \frac{\sigma_{xx}V_H}{V_x} \approx R_H\sigma_{xx}^2$ and of the change in conductivity $\Delta\sigma_{xx}$ on the incident photon energy $E$ are shown. While $\Delta\sigma_{xx}$ remains large and keeps increasing with increased photon energy beyond the A and B resonances (due to an enhancement in optical absorption), the anomalous Hall conductivity $\sigma_H$ peaks near the A feature and decreases quickly to almost zero at higher photon energies. Our observation is consistent with recent optical results indicating poor injection of valley polarization under off-resonance excitation due to the rapid intervalley relaxation of high-energy excited carriers *(10, 25)*.

Our experimental observation of a finite AHE only in monolayer MoS$_2$ under on-resonance, circularly polarized excitation strongly supports our interpretation of the signal as originating from the VHE. While a net spin polarization could also give rise to a finite AHE, the effect observed in our monolayer MoS$_2$ devices is mainly driven by a net valley polarization for the following two reasons. First, the majority carriers responsible for photoconduction are electrons, whose contribution to the AHE includes a negligible spin-polarized current due to the fast spin relaxations in the nearly spin-degenerate conduction band *(10, 27)*. Second, the coupling constant in the Hamiltonian responsible for the VHE *(6)* $\lambda_{VH} \sim a^2$ is much larger than that for the SHE *(28)* $\lambda_{SH} \sim \frac{\Delta_{SO}}{E_g} a^2 \sim 0.1 a^2$, where $a = 3.2$ Å and $\Delta_{SO} = 0.16$ eV are the lattice constant and the spin-orbit splitting in monolayer MoS$_2$ *(6)*, respectively.

In order to compare our results with the theoretical prediction in equation 1, we study the laser intensity (*P*) dependence of the photoinduced AHE under 657 nm excitation. For this, we first measure the gate dependence of $\Delta\sigma_{xx}$ at different incident laser powers (see Supplementary Materials for details). The effective change in the photoexcited carrier density $\Delta n_{ph}$ can then be estimated from the relation $\Delta\sigma_{xx} = \Delta n_{ph} e\mu$, where $\mu = \frac{1}{C_g}\frac{d\sigma_{xx}}{dV_g}$ is extracted from the dark electrical measurements (see figure 1C) with $C_g = 1.2\times10^{-8}$ F cm$^{-2}$ the back gate capacitance of our device. The quantity $\Delta n_{ph}$ should be equivalent to $\Delta n_v$ if the change in conductivity $\Delta\sigma_{xx}$ is solely driven by the valley-polarized carriers that are directly excited by resonant, circularly polarized light. In reality, however, $\Delta n_{ph}$ may include contributions from both valley-polarized and -unpolarized carriers; therefore, $\Delta n_{ph}$ provides an upper bound for $\Delta n_v$. The anomalous Hall conductivities for different gate voltages $V_g$ are shown as functions of $\Delta n_{ph}$ in figure 3. We also show the theoretical result predicted by equation 1 in the limit $\Delta n_{ph} = \Delta n_v$ in the same figure. For all gate voltages, $\sigma_H$ increases linearly with $\Delta n_{ph}$, consistent with the theoretical prediction. The anomalous Hall conductivity $\sigma_H$ also has the right order of magnitude and approaches the theoretical value at high $V_g$.

In the simplest case of $\Delta n_{ph} = \Delta n_v$, the effect should be independent of $V_g$, which is different from our experimental observations. One possible explanation of the discrepancy is the presence of photoconduction mechanisms that do not contribute to the AHE. Such mechanisms include the relaxation of valley polarization in a portion of the photoexcited carriers and the trapping of minority carriers whose effect is equivalent to electrostatic doping. Since the strength of disorder decreases when the device becomes more metallic at higher n-doping (see Supplementary Materials for $\rho_{xx}$ versus $V_g$ at different temperatures), a higher portion of photoexcited carriers could maintain the valley polarization and contribute to the Hall effect. We note, however, that the slope of the $\sigma_H$ vs $\Delta n_{ph}$ curves keeps increasing with higher $V_g$ and may go beyond the theoretical prediction. Unfortunately, the range of $V_g$ applied in our experiment is limited by the breakdown of the back gate, so we were unable to explore this regime. A second possibility is the changing relative importance of the intrinsic, side-jump and skew-scattering contributions *(21, 25)* to the VHE (note that the side-jump contribution is opposite in sign from the other two). The relative importance of each depends on the sample quality (e.g. the doping density and the amount of disorder). Studies of the dependence on temperature and on disorder are therefore required to better understand the doping density dependence of the VHE. Furthermore, a more accurate determination of $\sigma_H$ that takes into account the fringe fields in our Hall bar device may be needed for a better quantitative comparison.

In conclusion we have presented the first observation of the VHE in monolayer MoS$_2$. Our demonstration of the coupling between the electronic motion and the valley DOF in a 2D semiconductor and its sensitivity to photon polarization represents an important advance for both fundamental condensed matter physics and the emerging area of valley-dependent electronics.

**Acknowledgments:** We thank Daniel C. Ralph for his insightful suggestions and Joshua W. Kevek for technical support. We also thank Jie Shan for many fruitful discussions and Yumeng

You for private communications regarding the optical data on monolayer $MoS_2$. This research was supported by the Kavli Institute at Cornell for Nanoscale Science through the Cornell Center for Materials Research (NSF DMR-1120296). Additional funding was provided by the Nano-Material Technology Development Program through the National Research Foundation of Korea funded by the Ministry of Science, ICT and Future Planning (2012M3A7B4049887). Device fabrication was performed at the Cornell NanoScale Facility, a member of the National Nanotechnology Infrastructure Network, which is supported by the National Science Foundation (Grant ECCS-0335765). KLM acknowledges support from the NSF IGERT program (DGE-0654193) and the NSF GRFP (DGE-1144153).

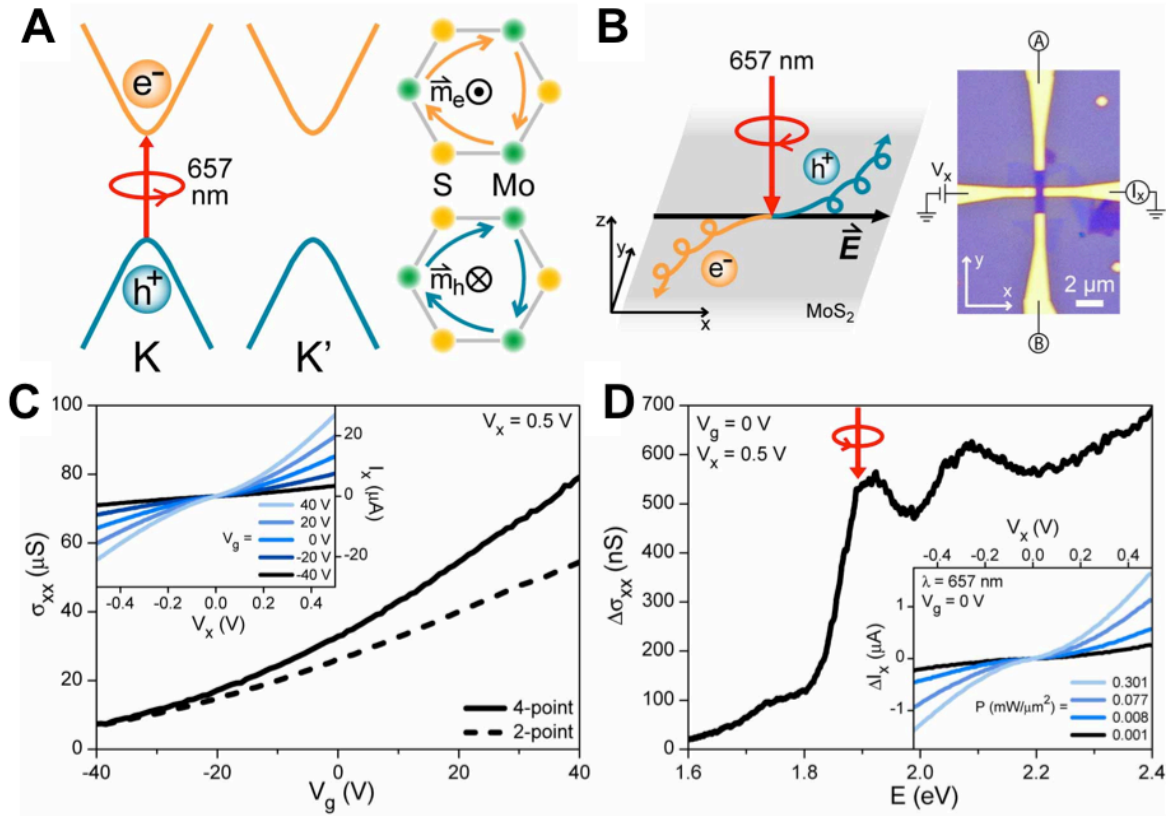

**Fig. 1 Monolayer MoS$_2$ Hall bar device.** **(A)** Schematics of the valley-dependent optical selection rules and the photoexcited carriers at the K valley that experience an effective magnetic field. **(B)** Schematic of a photoinduced AHE driven by a net valley polarization, and an image of the Hall bar device. **(C)** 2-point (dashed, $V_x$ = 0.5 V) and 4-point (solid) conductivities of the device as a function of back gate voltage $V_g$. Inset: Source-drain bias ($V_x$) dependence of the current along the longitudinal channel ($I_x$) at different back gate voltages $V_g$. **(D)** The change in conductivity $\Delta\sigma_{xx}$ as a function of incident photon energy E under laser illumination. The arrow indicates the excitation energy used in this experiment, E ≈ 1.89 eV. Inset: Source-drain bias ($V_x$) dependence of the photocurrent ($\Delta I_x$) at different incident laser intensities $P$ centered at 657 nm ($V_g$ = 0 V).

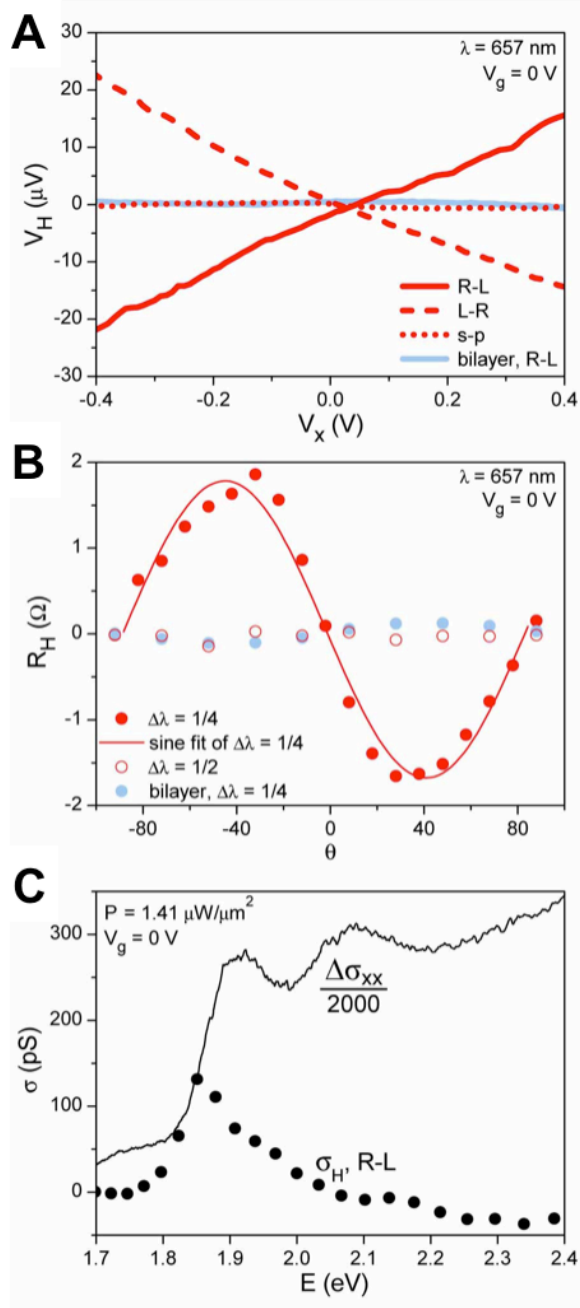

**Fig. 2 The valley Hall effect. (A)** The source-drain bias ($V_x$) dependence of the Hall voltage ($V_H$) for 657 nm R-L (red, solid) and L-R (red, dashed) modulations. Results from the monolayer device under half-wave (s-p) modulation (red, dotted) and from the bilayer device under R-L modulation (blue, solid) are also shown. **(B)** The anomalous Hall resistance of the monolayer device as a function of the incidence angle $\theta$ under quarter-wave ($\Delta\lambda = 1/4$, solid red) and half-wave ($\Delta\lambda = 1/2$, empty red) modulations. That of the bilayer device under quarter-wave modulation is also shown (blue). **(C)** Energy dependence of the change in conductivity $\Delta\sigma_{xx}$ (black curve) and of the anomalous Hall conductivity $\sigma_H$ (solid dots). The latter is obtained under R-L modulation.

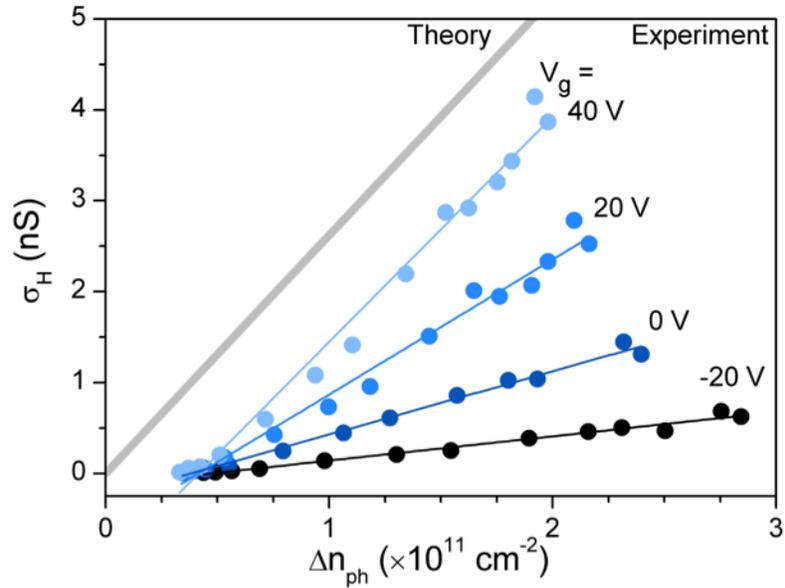

**Fig. 3 Doping dependence of the anomalous Hall conductivity.** The anomalous Hall conductivity as a function of the charge carrier density $\Delta n_{ph}$ at different gate voltages with linear fits to the experimental data. The theoretical prediction in equation 1, for $\Delta n_v = \Delta n_{ph}$, is shown by the grey curve.

**Supplementary Materials:**

Materials and Methods

Supplementary Text

Figures S1-S5

References (*S1-S9*)

**Supplementary Materials:**

# The Valley Hall Effect in MoS$_2$ Transistors

K. F. Mak[1,2]†, K. L. McGill[2], J. Park[1,3], and P. L. McEuen[1,2]*

Correspondence to: *mceuen@ccmr.cornell.edu, †km627@cornell.edu

## 1. Materials and Methods:

### 1.1 Device fabrication.

MoS$_2$ monolayers were mechanically exfoliated from bulk MoS$_2$ crystals onto Si substrates coated by 300 nm of SiO$_2$. Monolayer samples were identified using a combination of optical contrast and photoluminescence spectroscopy *(S1)*. Standard electron beam lithography techniques were used to define metal contact areas on our exfoliated samples. Electron beam evaporation was used to deposit 0.5 nm Ti/50 nm Au contacts, followed by a standard methylene chloride/acetone lift-off procedure. Using electron beam lithography to create an etch mask, we defined the Hall bar geometry using a ten-second low-pressure SF$_6$ plasma etch. Finally, the device was laser annealed in high vacuum *(S2, S3)* (~ 10$^{-6}$ torr) at 120 $^0$C for ~10 hours before measurement. We note that the reasons for creating a Hall bar device with a long Hall probe and a short photoconduction channel (figure 1B in main text) are two-fold: 1) we want the photocurrent (which is generated most efficiently at the contacts) to be produced near the center of the device so that any Hall voltage can be efficiently picked up by the Hall probe; 2) we want to reduce the background photovoltage generated at the metal-semiconductor contacts of the Hall probe (see below).

### 1.2 Photoconduction and Hall voltage measurements.

Measurements were performed in a Janis cryostat cooled by liquid nitrogen and placed on an inverted microscope. A standard Hall voltage measurement was performed with a source-drain voltage $V_x$ applied across the short channel as shown in figure 1B in the main text. The voltage difference between the A and B contacts of the Hall probe was measured by a voltage amplifier, whose output was further sent to a lock-in amplifier. For our photocurrent measurement, a Fianium supercontinuum laser source with a monochromator (selecting a line width of ~5 nm for each color) was used for acquiring the photoconductivity and Hall conductivity spectra. Diode lasers (657 nm) were used for all other optical excitations. To modulate the polarization of the incident light, the laser was linearly polarized and passed through a photoelastic modulator before being focused onto the sample through a 40x long working distance objective (spot diameter between 1–3 μm depending on the specific measurement). The angle of incidence $\theta$ of the linearly polarized light with respect to the fast axis of the modulator was varied by a half waveplate so that the photon ellipticity could be continuously tuned while being modulated at 50 kHz. For the control experiments involving modulation with linearly polarized light, the phase shift of the modulator was switched from quarter-wave ($\Delta\lambda = 1/4$) to half-wave ($\Delta\lambda = 1/2$) modulation. Photocurrent and Hall voltage maps were obtained by scanning the laser spot across the samples with a pair of scanning mirrors, and reflection images were obtained by collecting the reflected light in a silicon photodiode.

## 2. Supplementary Text:

### 2.1 Derivation of the anomalous Hall conductivity

According to ref. S4 and S5, the Hall conductivity for the electrons in the K' valley of an MoS$_2$ monolayer originating from the intrinsic Berry curvature effect can be written as

$$\sigma_{H,K'} = \frac{\pi e^2}{h} \int_{E_g/2}^{\infty} d\epsilon\, g(\epsilon) \Omega_{e,K'}(\epsilon) f_e(\epsilon). \tag{S1}$$

Ignoring spin-orbit coupling, $g(\epsilon) = \frac{2m_e \epsilon}{\pi \hbar^2 E_g}$ is the electron density of states at the K' valley, $\Omega_{e,K'}(\epsilon) = \frac{\hbar^2 E_g^2/m_e}{8\epsilon^3}$ is the Berry curvature and $f_e(\epsilon)$ is the Fermi-Dirac distribution. In the degenerate limit, $\sigma_{H,K'}$ becomes

$$\sigma_{H,K'} \approx \frac{e^2}{h} \frac{\mu_{K'}}{E_g} = \frac{e^2}{h} \frac{\hbar^2 \pi n_{K'}}{2 m_e E_g}, \tag{S2}$$

where $\mu_{K'}$ is the chemical potential and $n_{K'}$ is the total electron density at the K' point. In this limit, the anomalous Hall conductivity $\sigma_H$ becomes (including only the electron contribution)

$$\sigma_H \approx \frac{e^2}{h} \frac{\hbar^2 \pi \Delta n_v}{2 m_e E_g}. \tag{S3}$$

Here, $\Delta n_v$ is the carrier density imbalance between the two valleys generated by photoexcitation and $m_e$ is the electron band mass.

In the nondegenerate limit, we can show that $\sigma_{H,K'}$ becomes

$$\sigma_{H,K'} \approx \frac{e^2}{h} \frac{\hbar^2 \pi n_{K'}}{2 m_e E_g} F\left(\frac{E_g}{2 k_B T}\right), \tag{S4}$$

where $F \approx 1$ is dimensionless and is weakly dependent on temperature. Equation S4 thus reduces to equation S3, so the expression for the Hall conductivity $\sigma_H$ when expressed in terms of the carrier density is approximately the same in both the degenerate and nondegenerate limits. The above derivation is for the intrinsic Berry curvature effect. It is shown in ref. S5 that the side-jump contribution is twice as big as the intrinsic effect and has the opposite sign. Thus, including the intrinsic and side-jump contributions, the anomalous Hall conductivity can be reduced to equation 1 in the main text.

### 2.2 Temperature-dependent electrical transport

To better understand the electrical transport properties of our device, we show the temperature dependence of the resistivity $\rho_{xx}$ versus gate voltage $V_g$ in figure S1A. We clearly see the presence of a metal-insulator transition across $V_g = 0$ V: the resistivity increases with decreasing temperature for $V_g < 0$ V (the insulating regime) and vice versa for $V_g > 0$ V (the metallic regime). This is further illustrated in figure S1B, which shows the temperature dependence of the resistivity at different $V_g$. Consistent with recent observations *(S2, S6)* and with previous studies on 2D electron gases in various semiconductor systems *(S7)*, the transition occurs near a resistivity value of $\rho_{xx} \sim \frac{h}{e^2} = 2.6 \times 10^4 \, \Omega$ that obeys the Ioffe-Regel criterion *(S8)* $k_F l \sim 1$. Here $k_F$ and $l$ are the Fermi wave-vector and the mean free path of the electrons, respectively.

## 2.3 Scanning Hall voltage microscopy

We characterized our device under illumination by spatially mapping its photocurrent and Hall voltage responses. All the maps were recorded at $V_g = 0$ V, using a 647 nm continuous wave laser (spot diameter ~1 μm) at an incident power of ~50 μW. Figure S2A shows the scanning photocurrent image of a second monolayer device at a bias voltage of $V_x = 0.5$ V. The photocurrent is mainly generated at the center of the device where a source-drain bias voltage is applied across the short channel. The corresponding scanning Hall voltage ($V_H$) images are shown in figures S2B and C for R-L and L-R modulations, respectively. We see that a finite Hall voltage is produced at the center of the device, coinciding with the location of photocurrent production. Furthermore, the sign of $V_H$ reverses when the helicity of the modulation changes from R-L to L-R.

In figure S2D, E and F, we show the results from a bilayer device control experiment. Although a similar photocurrent is again produced at the center of the device, the Hall voltage is much smaller (by about a factor of 10) than that of the monolayer device. We note that significant photovoltages (particularly in the bilayer device) are also observed at the metal-semiconductor contacts of the Hall probe (both at zero and finite bias along the short channel). These photovoltages probably arise from the modification of the polarization state by the metal contacts, which leads to a corresponding power modulation. The presence of such undesirable signals is the reason we use a long Hall probe in our experiment. Overall, the observation of a significant Hall voltage only at the center of monolayer devices confirms the observation of the VHE.

## 2.4 Data from extra monolayer devices

Figure S3A shows a Hall voltage ($V_H$) measurement from a second monolayer device. Again, we see a finite Hall voltage that scales linearly with the source-drain voltage $V_x$ only for the L-R and R-L modulations of the incident laser beam. The Hall voltage vanishes under half-wave modulation. Figure S3B shows the sine dependence of the Hall resistance $R_H$ on the angle $\theta$ under quarter-wave modulation, which vanishes when switched to half-wave modulation. We have also observed these effects in three other monolayer devices (data not shown).

## 2.5 Photodoping density

We extract the photoexcited carrier density $\Delta n_{ph}$ by measuring the gate dependence of $\Delta \sigma_{xx}$ at different incident laser powers (inset of figure S4A). As mentioned in the main text, $\Delta n_{ph}$ can be obtained from the relation $\Delta \sigma_{xx} = \Delta n_{ph} e \mu$ with $\mu = \frac{1}{C_g} \frac{d\sigma_{xx}}{dV_g}$. The photoexcited carrier density $\Delta n_{ph}$ as a function of back gate voltage $V_g$ at different excitation intensities is shown in figure S4A. At high excitation intensities, a charge density $\Delta n_{ph}$ on the order of $10^{11}$ cm$^{-2}$ is seen.

Figure S4B shows the dependence of the carrier density $\Delta n_{ph}$ on the incident laser intensity $P$ at different gate voltages. The observed saturation behavior might be explained by the presence of trapped-charge contributions to the photoconduction, as it is similar to the observed intensity saturation in disorder-induced photoluminescence that originates from the change in occupancy of the trapped states *(S9)*. More systematic studies of the dependence of the photoconduction on the amount of disorder in the system are required for a better understanding of the laser power dependence of the photoresponse.

**2.6 Gate voltage dependence of $\sigma_H$**

The gate voltage ($V_g$) dependence of the anomalous Hall conductivity $\sigma_H$ under R-L modulation (center wavelength 657 nm with an excitation intensity of 150 µW µm$^{-2}$) is shown in figure S5; note that it increases with electron doping. As mentioned in the main text, no dependence on the gate voltage is expected according to the simplest theoretical model. Possible explanations for this discrepancy have been discussed in the main text by considering the portion of $\Delta n_{ph}$ that contributes to the Hall effect and the presence of extrinsic contributions.

**Supplementary Figures:**

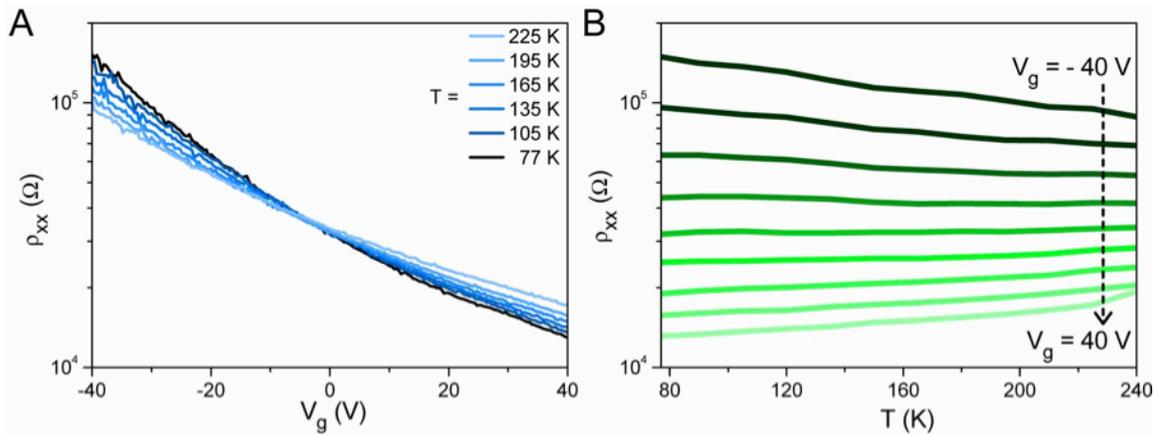

**Fig. S1 (A)** Resistivity $\rho_{xx}$ of the monolayer device as a function of back gate voltage at different temperatures. **(B)** Temperature dependence of $\rho_{xx}$ at different back gate voltages. A metal-insulator transition is observed near $\rho_{xx} \sim \frac{h}{e^2} = 2.6 \times 10^4 \, \Omega$.

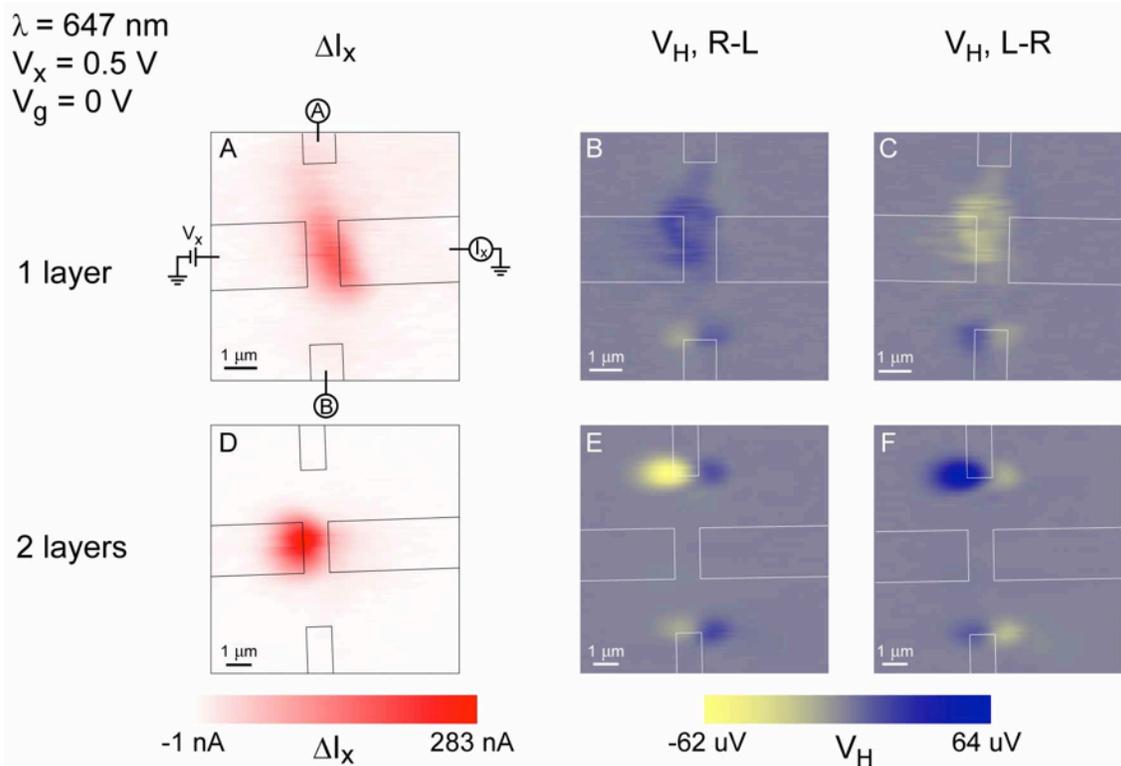

**Fig. S2 Scanning photocurrent and Hall voltage images.** The measurement schematic for all maps is indicated in **(A)**; all maps were recorded under 647 nm excitation at a power of ~50 μW and a spot diameter of ~1 μm. **(A)** Scanning photocurrent image of a monolayer device at a source-drain bias $V_x$ = 0.5 V ($V_g$ = 0 V). The corresponding scanning Hall voltage image under **(B)** R-L and **(C)** L-R modulations, respectively. **(D-F)** are the corresponding images of a bilayer device.

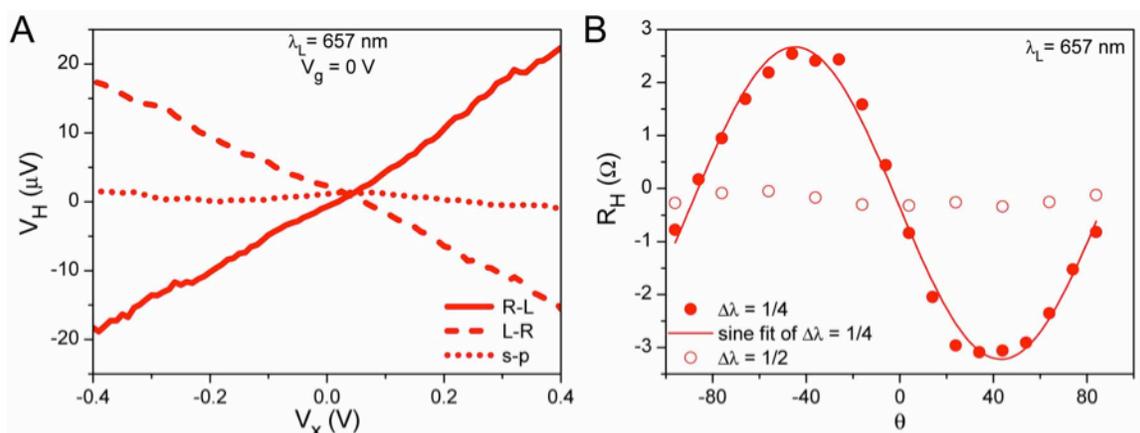

**Fig. S3 Extra data showing the VHE in a second monolayer device.** **(A)** The source-drain bias dependence of the Hall voltage for 657 nm R-L (red, solid) and L-R (red, dashed) modulations. The result for half-wave (s-p) modulation (red, dotted) is also shown. **(B)** The anomalous Hall resistance as a function of the incidence angle $\theta$ under quarter-wave ($\Delta\lambda$ = 1/4) and half-wave ($\Delta\lambda$ = 1/2) modulations.

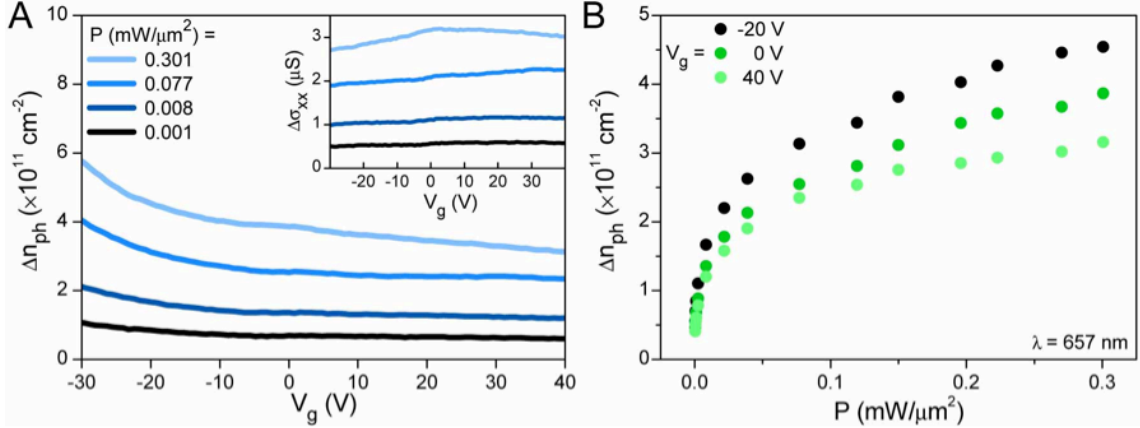

**Fig. S4 (A)** The photoexcited carrier density $\Delta n_{ph}$ as a function of gate voltage $V_g$ at different laser excitation intensities $P$. The inset shows the corresponding $V_g$ dependence of $\Delta\sigma_{xx}$ from which the carrier densities are extracted. **(B)** The carrier density $\Delta n_{ph}$ as a function of laser intensity $P$ at different gate voltages $V_g$.

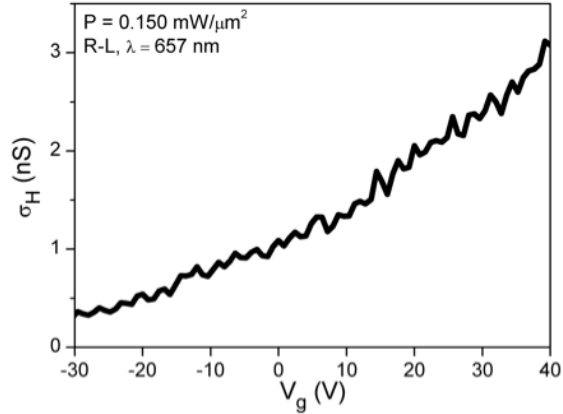

**Fig. S5** The $V_g$ dependence of the anomalous Hall conductivity $\sigma_H$ at a laser excitation intensity of ~ 150 µW µm$^{-2}$.